\documentclass[aps,prb,twocolumn,superscriptaddress,preprintnumbers,amsmath,amssymb]{revtex4}

\usepackage{graphicx}
\usepackage{dcolumn}
\usepackage{bm}
\newcommand{\cro}{Cr$_2$O$_4$}
\newcommand{\ir}{IR}
\newcommand{\wns}{cm$^{-1}$}
\begin{document}


\title{Spin phonon coupling in frustrated magnet CdCr$_2$O$_4$}

\author{R. Vald\'{e}s Aguilar}
\email{rvaldes@physics.umd.edu}
 \affiliation{MRSEC, University of Maryland, College Park, Maryland 20742}

\author{A. B. Sushkov}%
 \affiliation{MRSEC, University of Maryland, College Park, Maryland 20742}%
\author{Y. J. Choi}
\affiliation{Rutgers Center for Emergent Materials and Department of Physics \& Astronomy, Rutgers University, Piscataway, New Jersey 08854}%
\author{S-W. Cheong}
\affiliation{Rutgers Center for Emergent Materials and Department
of Physics \& Astronomy, Rutgers University, Piscataway, New
Jersey 08854}
\author{H. D. Drew}
\affiliation{MRSEC, University of Maryland, College Park, Maryland
20742}


\begin{abstract}
The infrared phonon spectrum of the spinel Cd\cro\ is measured as a
function temperature from 6 K to 300K.  The triply degenerate Cr
phonons soften in the paramagnetic phase as temperature is lowered
below 100 K and then split into a singlet and doublet in the low
temperature antiferromagnetic phase which is tetragonally distorted
to relieve the geometric frustration in the pyrochlore lattice of
Cr$^{3+}$ ions. The phonon splitting is inconsistent with the simple
increase (decrease) in the force constants due to decreasing
(increasing) bond lengths in the tetragonal phase. Rather they
correspond to changes in the force constants due to the magnetic
order in the antiferromagnetic state.  The phonon splitting in this
system is opposite of that observed earlier in Zn\cro\ as predicted
by theory. The magnitude of the splitting gives a measure of the
spin phonon coupling strength which is smaller than in the case of
Zn\cro.
\end{abstract}

\maketitle

\section{Introduction}
Due to geometric frustration antiferromagnetically coupled
Heisenberg spins in the pyrochlore lattice do not order at any
finite temperature \cite{Anderson-pyrochlore}. Geometric
frustration also has the consequence of creating very large
degeneracy in the ground state giving a finite entropy at zero
temperature. This degeneracy makes them susceptible to ordering
due to perturbations such as magnetoelastic
\cite{Tcherny-prl-string,Tcherny-prb-pyro} or further neighbor
exchange coupling, and the possibility of exotic ground states.
These features have led to great interest in this class of
materials \cite{Ramirez-geometric}. Their properties are being
extensively studied theoretically and experiments are dedicated to
look for materials that do not order, the so-called spin-ice
materials \cite{Ramirez-spin-ice}.

The family of spinel compounds \textit{A}\cro\ (\textit{A} = Zn, Cd
or Hg) is a good example of the Heiseinberg antiferromagnet in the
pyrochlore lattice, where the only magnetic ion is the non
Jahn-Teller active Cr$^{3+}$ (with spin S=3/2). The Cr ions sit at
the vertices of the corner sharing tetrahedra spanned by the lattice
with space group $Fd\overline{3}m$, this arrangement causes
geometric frustration of the primarily antiferromagnetic
interactions. It has been shown that magnetoelastic coupling leads
to a structural distortion \cite {Lee-dist-ZnCr,Lee-dyn-CdCr} in
both Zn\cro\ and Cd\cro\ and that in this distorted lattice a
complex antiferromagnetic order is established below $T_N$ = 12 and
7.8 K, respectively. The uniform component of the lattice
distortions is tetragonal with $(c-a)/a = 5\times10^{-3}$ and $=
-1.5\times10^{-3}$ for Cd\cro\ and Zn\cro, respectively. The
opposite sign of the observed distortion in these very similar
materials is unexpected, making their differences significant and
worthy of investigation. It has been also
observed\cite{Sushkov-ZnCr} that the distortion in Zn\cro\ leads to
splitting of one of the infrared (\ir) active phonons and that the
size of the splitting gives a measure of the spin correlations both
above and below $T_N$.

A Landau theory of the magnetoelastic interaction in the
pyrochlore lattice for the case of uniform distortions has been
developed by \citet{Tcherny-prl-string,Tcherny-prb-pyro}. They
found a correlation between the lattice distortions, or bond
order, and the spin order of the ground state.  Since it is known
that the lattice distortion in Zn\cro\ is not uniform
\cite{Lee-distortions}(i.e. $\mathbf{q} \neq 0$), this theory does
not apply directly. It is, nevertheless, a good starting point in
the explanation of some experimental facts and, as we will show
below, it allows an understanding of certain features of the
behavior of the \ir\ phonons in Cd\cro\ as well. It is
particularly noteworthy that even though the structural
differences between Zn\cro\ and Cd\cro\ are minimal at high
temperatures \cite{Lee-dist-ZnCr,Lee-dyn-CdCr}, significant
differences appear in the \ir\ spectra of the two upon magnetic
ordering, suggesting that the magnetic interactions are very
sensitive to subtle lattice changes.

It is important therefore to compare these two materials in order to
elucidate how the subtle differences in the radius of the $A$ site
ion leads to characteristically different distortions and phonon
splittings in the ground state.  In this paper we report the
temperature dependence of the \ir\ reflectivity spectra in the
phonon frequency range of the frustrated antiferromagnet Cd\cro. We
find that only one of the triply degenerate modes in this compound
splits significantly below the N\'{e}el temperature, similar to the
splitting in Zn\cro [\onlinecite{Sushkov-ZnCr}]. However, some important
differences are observed. We discuss these effects in terms of the
lattice distortion that relieves frustration and its associated spin
configuration.

\section{Results}
Single crystals were grown by a flux method as described elsewhere
\cite{Dabkowska-flux}. Large surfaces of the (111) plane were
polished for reflection measurements, typical sizes were
3$\times$3$\times$0.5 mm$^3$. The samples were characterized by
magnetic susceptibility measurements and showed the
antiferromagnetic transition temperature to be $T_N$ = 7.8 K. The
temperature dependence from 6 to 300 K of the reflectivity spectra
($\hbar\omega = 15$--$100$ meV) was obtained using a Fourier
transform spectrometer with the sample in vacuum in an optical
cryostat with continuous He flow for cooling
\cite{Sushkov-ZnCr,Rolando-TbMn2O5}. We fitted \cite{Reffit} the
spectra with a sum of Lorentzians model for the dielectric constant\
\begin{equation}
\varepsilon(\omega) = \varepsilon_\infty+\sum_j
\frac{S_j}{\omega_{j}^2-\omega^2-i\omega\gamma_j}
\end{equation}
using the reflectivity formula
$R=|\sqrt{\varepsilon}-1|^2/|\sqrt{\varepsilon}+1|^2$. Then we
extracted the parameters of the Lorentzian oscillators as a
function of temperature: $S_j$ (cm$^{-2}$) spectral weight,
$\omega_j$ (cm$^{-1}$) phonon frequency and $\gamma_j$ (cm$^{-1}$)
the linewidth.

\begin{figure}
\includegraphics[width=.9\columnwidth]{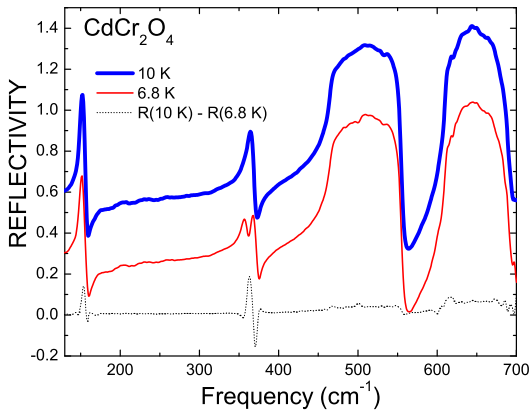}
\caption{(Color online) 10 K (thick line) and 6.8 K (thin line)
reflectivity spectra of Cd\cro (offset for clarity). Also shown is
the difference spectra in thin dotted line.} \label{spectra}
\end{figure}

In the paramagnetic phase the spectra contains only the four triply
degenerate \ir\ modes allowed by symmetry (4$T_{1u}$). Below $T_N$
five modes are observed as seen in figure \ref{spectra}. In the
ordered phase, due to the tetragonal distortion, the phonon triplets
should split into doublets and singlets \cite{Fennie-LSDA}, but only
the phonon at 365 cm$^{-1}$ splits significantly below $T_N$. This
phonon mode contains the largest component of the symmetry
coordinate \cite{Himmrich-ZnCr} that modulates the Cr-Cr distance
and, thus, the direct exchange between neighbors. Therefore we
expect the effects of the transition into the antiferromagnetic
state to be much more pronounced on this mode. We note, however,
that by taking the difference between spectra above and  below $T_N$
we see that the low frequency ($\sim 150$ \wns) phonon also shows
signature of splitting. We also point out that the motion involved
in the two high frequency phonons modulates the superexchange path
Cr-O-Cr but we do not observe any splitting on these phonons,
even though this phonons also contain a small component of the Cr-Cr
mode that dominates the low frequency phonons. We can
conclude then that the main component of the magnetic interaction is
the direct exchange between Cr ions.

\begin{figure}
\includegraphics[width=.9\columnwidth]{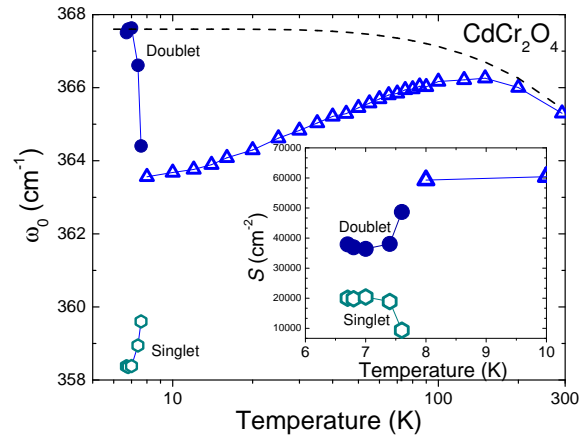}
\caption{(Color online) Temperature dependence of the Cr phonon
frequency. The dashed line indicates the anharmonic behavior of the
frequency. The inset shows the spectral weight of this phonon and
its distribution below $T_N$.} \label{freq}
\end{figure}

Above $\sim$ 150 K the behavior of the frequency of the Cr phonon,
shown in the main panel of figure \ref{freq}, is as expected due to
anharmonic processes, but below this temperature the phonon softens
by 3 cm$^{-1}$. As we will discuss below, we understand this effect
as a signature of the coupling of this phonon to the spin
fluctuations in the paramagnetic phase and it is similar to the
behavior in Zn\cro. Below $T_N$ the phonon splits into a doublet and
a singlet. The singlet frequency softens significantly and the
doublet frequency hardens.  The low temperature limit of the doublet
frequency is close to the expected value in the paramagnetic state
in the presence of anharmonic hardening of the lattice, but with the
absence of softening due to spin liquid effects as indicated by the
dashed line in figure \ref{freq}. By comparing their spectral
weights, we conclude that the singlet shifts down and the doublet
shifts up in frequency upon cooling below $T_N$, with a final
splitting $\Delta =$ 9 cm$^{-1}$. This value of the splitting is
close to $\Delta_{ZCO}$ = 11 cm$^{-1}$ reported
before\cite{Sushkov-ZnCr} in Zn\cro, where the doublet softens and
the singlet hardens. The different behaviors of this phonon in
Zn\cro\ and Cd\cro\ will be explained as a consequence of the spin
phonon coupling effects on their different magnetic order.

\begin{figure}
\includegraphics[width=.9\columnwidth]{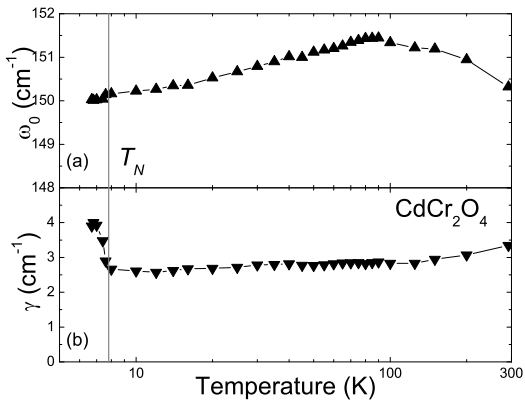}
\caption{Temperature dependence of the lowest energy phonon (a)
frequency and (b) linewidth.} \label{3-phon}
\end{figure}

Figure \ref{3-phon}(a) shows the temperature dependence of the
frequency of the lowest energy \ir\ active phonon. This phonon
also shows anomalous softening below 100 K, the frequency shifts
by 1.5 \wns\ from above $T_N$ to approximately 100 K. This
temperature dependence is similar to the behavior of the Cr
phonon triplet at 365 \wns, therefore we suspect that the origin
of the this behavior is the same; coupling of these phonons
to the antiferromagnetic fluctuations in the paramagnetic phase.
Even though we could not fit the lineshape of this phonon with two
separate oscillators significantly better than with a single
oscillator, it is evident from the sudden increase of the
linewidth at $T_N$ (figure \ref{3-phon}(b)) that this phonon also
splits. The high frequency phonons show the expected increase on
cooling of its frequency due to anharmonic processes and do not
show any evidence of splitting below $T_N$.

\section{Discussion}
The lattice distortions that effectively couple to the spins, and
release the frustration, belong to the $E$ (doublet)
representation of the point group of the tetrahedron $T_d$. These
distortions correspond to tetragonal and orthorhombic
modifications of the unit cell. When the full translational
symmetry of the lattice is taken into account, the $E$
representation becomes $E_u$ and $E_g$ according to whether the
distortion is odd or even under inversion symmetry. The $E_g$
distortion is then a uniform stretching of all tetrahedra, while
the $E_u$ distortion stagers stretching and contraction along one
axis in neighboring tetrahedra. Each of these distortions is
accompanied by different spin configurations, and are displayed in
figures 5 and 6 of \citet{Tcherny-prb-pyro}.

\begin{figure}[t]
\includegraphics[width=.8\columnwidth]{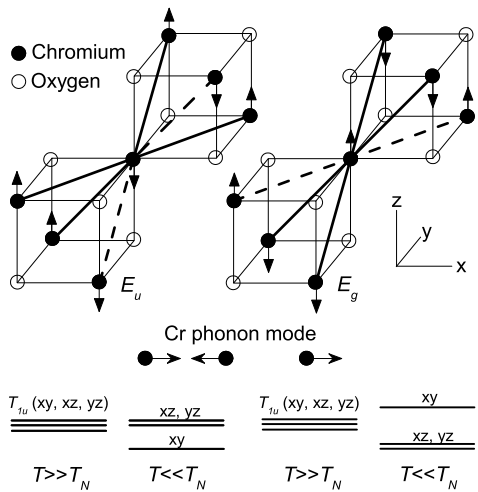}
\caption{Expected spin order effect on the frequencies of the Cr
phonon triplet. (Top) Spin and bond orders induced by the $E_u$
(left) and $E_g$ (right) distortions, solid (dashed) lines represent
satisfied (frustrated) bonds. (Bottom) Schematic representation of
the phonon triplet splitting induced by the spin ordering
corresponding to each distortion.} \label{distortion}
\end{figure}

Based on this model we can qualitatively explain the phonon
splitting and how the spectral weight is redistributed below $T_N$.
\citet{Chern-CdCr} proposed a model to explain the
observed\cite{Lee-dyn-CdCr,Lee-spiral-CdCr2O4} spin configuration in
Cd\cro, wherein the major contribution to the lattice distortion
comes from the phonon with $E_u$ symmetry with a smaller
contribution from the $E_g$ phonon. The distortion is such that the
contraction occurs along the $x$ axis in one tetrahedron and in the
$y$ axis in the neighboring one. The bonds along [110] and
[1$\overline{1}$0] are fully satisfied (i.e. neighboring spins are
always antiparallel), whereas the bonds along [011],
[0$\overline{1}$1], [101] and [$\overline{1}$01] alternate between
frustrated and satisfied bonds as shown in the left side of figure
\ref{distortion} with full lines for satisfied bonds and dashed
lines for frustrated bonds. These bonds are formed along chains of
up-up-down-down spins around which one can form left and right
handed screws of fully frustrated and fully satisfied bonds.

In this case the phonon triplet $T_{1u}$ mode (motion indicated in
the middle panel of figure \ref{distortion}) involves different
bonds in each of its components: the $xy$ motion probes the fully
satisfied bonds, and the $yz$ and $xz$ motions probe the mixed
bonds, as illustrated at the top of figure \ref{distortion}. This model serves
as a natural explanation to the splitting of the phonon frequencies,
the doublet ($yz$ and $xz$) should not change its frequency with
respect to the paramagnetic phonon triplet since the contribution
from frustrated and satisfied bonds would cancel (i.e. $\langle
S_i\cdot S_j\rangle =$ 0), while the phonon singlet would reduce its
frequency since the bond is fully satisfied ($\langle S_i\cdot
S_j\rangle = -1$). This picture is illustrated by the results in
figure \ref{freq}. These considerations are consistent with
observation when the spin glass phonon softening effects are
included. Therefore, in the N\'{e}el phase the phonon doublet shifts
to the position it would have in the absence of any spin
correlations at low temperatures.

This simple explanation also helps us understand the results
obtained in Zn\cro. In this compound the observed uniform distortion
is given by a simple lattice contraction along the [001] direction.
So the $E_g$ distortion is dominant. The bond order is such that
[110] and [1$\overline{1}$0] bonds are fully frustrated, and [011],
[0$\overline{1}$1], [101] and [$\overline{1}$01] bonds are always
satisfied. The doublet then ($xz$ and $yz$) has lower frequency than
the singlet ($xy$) as observed in the \ir\ measurements
\cite{Sushkov-ZnCr} and indicated on the right side of figure
\ref{distortion}.

If we compare the split phonon frequencies with the phonon triplet
just above $T_N$, and not with the paramagnetic state as done above,
we expect the frequency shifts to have a definite behavior. Just
above the transition individual tetrahedra cycle through different
spin configurations, that we assume to be collinear, so that each
bond spends 1/3 of the time with parallel spins and 2/3 of it with
antiparallel spins. Then on average we will have $\langle S_i\cdot
S_j\rangle = -1/3$, which means that the phonon doublet shifts up
1/3$\Delta$ and the singlet shifts down 2/3$\Delta$ with respect to
the frequency just above $T_N$. The experimental observation does
not agree with this prediction. This could be an indication of the
role that the non-collinear states play near the phase transition that is not captured by the model
\cite{Tcherny-prl-string,Tcherny-prb-pyro} we have used.

It is also important to note that the observation in Cd\cro\
(Zn\cro) is in opposition to what is expected from a simple
tetragonal distortion that elongates (contracts) the $c$ axis when
the distortion does not couple to the spins. For a simple
elongation along the $c$ axis (as it would correspond to Cd\cro)
we expect the frequency of the singlet ($xy$ motion) to go up
since the effective force constant of the bond is increased in
proportion to the distortion, and the doublet ($xz$, $yz$ motions)
would have lower energy. For the contraction case (as in Zn\cro)
the force constant diminishes, making the singlet frequency to go
down and the doublet up. This makes clear the crucial role in
relieving frustration that the spin-phonon coupling plays in these
materials. We summarize the experimental observation in figure
\ref{distortion} where the splitting of the phonon triplet is
shown schematically.

The behavior of the phonon frequencies of the two lowest energy
phonons in Cd\cro, softening above $T_N$ on cooling, is similar to
the Cr phonon triplet reported\cite{Sushkov-ZnCr} for Zn\cro. This
effect could be explained as the consequence of spin-phonon coupling
due to the short range magnetic order in the spin glass state at low
temperatures and how it affects the phonon frequency. Since direct
exchange between Cr ions dominates the magnetic interactions, the
dependence of the exchange on bond separation induced by the spin
correlations modifies the phonon frequency with a term proportional
to the nearest neighbor spin-spin correlation function,
$\omega=\omega_0 + \lambda\langle S_i\cdot S_j\rangle$, where
$\lambda$ is proportional to the second derivative of the direct
exchange constant with respect to the phonon coordinate. Even though
there is no magnetic long range order above $T_N$, the spin-spin
correlation function is not zero. Its value could be estimated from
the magnetic specific heat\cite{Sushkov-ZnCr}. Unfortunately there
are no specific heat data for Cd\cro\ that would allow a direct
comparison of the value of the spin phonon constant $\lambda$ among
the chromates. We can also estimate the value of $\lambda$ from the
doublet-singlet splitting\cite{Fennie-LSDA} $\Delta =$ 9 \wns.
Using the spin-Peierls order parameter, $\langle S_1\cdot S_2 -
S_2\cdot S_3\rangle =$ 9/4, we obtain $\lambda =$ 4 \wns\, which is
somewhat smaller than the value in Zn\cro\ obtained
before\cite{Sushkov-ZnCr,Fennie-LSDA}. The value of the ratio of
$\lambda_{Zn}/\lambda_{Cd}$ ($\approx 1.3$) is much smaller than the expected from
the ratio of the total exchange constants $J_{Zn}/J_{Cd} \approx 4$
as obtained from the Curie-Weiss fit to susceptibility data.

This apparent discrepancy is likely a consequence of the oversimplification of our
representation of the magnetic order in these materials. The value of the
spin-Peierls order parameter used for the estimation of $\lambda$ was obtained
for the collinear configurations of reference [\onlinecite{Tcherny-prb-pyro}]. Since the magnetic structure
of Zn\cro\ is noncollinear\cite{Lee-struct-ZnCr}, we expect that the spin-Peierls order
parameter be much smaller than the value used here, which would make the value of $\lambda_{Zn}$
and the ratio $\lambda_{Zn}/\lambda_{Cd}$ correspondingly larger. A comparison of our estimate
of $\lambda$ with the value obtained from the magnetic specific heat as done earlier for
Zn\cro\ [\onlinecite{Sushkov-ZnCr}] would provide a test of these ideas.
Therefore, the measurement of magnetic specific heat in Cd\cro\ is a priority.

The effects of the competition between direct exchange and
superexchange in the spin-lattice interaction in the magnetic
chromium spinels has been recently studied using \ir\ spectroscopy
by \citet{Rudolf-spinels}. Their results demonstrate that the
simple picture presented here needs modification if it is to be
applied to systems where the direct antiferromagnetic exchange is
not the dominant interaction. We note, however, that their
measurements in Cd\cro\ do not agree with ours completely, and we
believe that their use of polycrystalline samples might be a
factor for the differences.

\section{Conclusions}
We have presented measurements of the \ir\ phonon spectra of
frustrated antiferromagnet Cd\cro. Based on the model of the soft
pyrochlore lattice \cite{Tcherny-prl-string,Tcherny-prb-pyro} we
have explained the behavior of the frequency in the
Cr-motion-dominated phonon triplet as it enters into the
antiferromagnetic phase. This model also allows an understanding of
the different spectral weight distributions between Cd\cro\ and
Zn\cro. We found as well the value of the spin-phonon coupling
constant to be smaller in Cd\cro\ than in Zn\cro. The comparison
between these two materials demonstrates our understanding that
direct exchange is the most relevant interaction for these systems
and of the main features of the interplay between frustrated
magnetism and spin-lattice coupling in the pyrochlore lattice.

\section{Acknowledgements}
We thank O. Tchernyshyov, G-W. Chern and C. Fennie for useful
discussions. This work was supported in part by the National
Science Foundation MRSEC grant DMR-0520471.

\bibliography{cdcr2o4}
\end{document}